\documentclass[a4paper,fleqn,usenatbib]{mnras}

\usepackage[T1]{fontenc}
\usepackage{ae,aecompl}
\usepackage[normalem]{ulem}

\usepackage{graphicx}	
\usepackage{amsmath}	
\usepackage{amssymb}
\usepackage{subcaption}
\usepackage{url}
\title[CIR and hot gas properties of ETGs]{Connections between Central Intensity Ratio and hot gas properties of early-type galaxies}

\author[Vinod et al.]{
K T Vinod,$^{1}$\thanks{E-mail: vinod2085@gmail.com}
C Baheeja,$^{1}$
and C D Ravikumar$^{1}$
\\
$^{1}$Department
of Physics, University of Calicut, Malappuram-673635, India\\
}

\date{Accepted XXX. Received YYY; in original form ZZZ}

\pubyear{2022}
\newcommand{\lx}{\rm L$_{\rm X,GAS}$}
\begin{document}
\label{firstpage}
\pagerange{\pageref{firstpage}--\pageref{lastpage}}
\maketitle


\begin{abstract}
We report strong connections between central intensity ratio (CIR) and hot gas properties of Early-type galaxies (ETGs) in the nearby ($\rm D<30\, Mpc$) Universe. We find new strong correlations between (optical) CIR and X-ray gas luminosity ($\rm L_{\rm X,GAS}$) as well as X-ray gas temperature ($\rm T_{GAS}$). These correlations suggest that higher the central gas temperature lower will be the (central) star formation process in ETGs. Correlations of CIR separately with K-band magnitude and age of the sample galaxies, further support suppression of star formation in the central region of ETGs as they grow in mass and age. The systematic and tight variation of CIR with $\rm L_{\rm X,GAS}$ not only shows its remarkable potential to estimate $\rm L_{\rm X,GAS}$ from simple photometry but also helps in transforming the core-coreless dichotomy into a gradual one.
\end{abstract}




\section{Introduction}

The evolution scenario of galaxies is believed to be migrated from blue to red sequence via the quenching of star-formation. But, the driving mechanism behind this migration process is still unclear. Early-type galaxies (ETGs) possess a vast amount of hot gas at their interstellar medium (ISM) during this migration process, and this hot, diffuse gas is one of the prime sources of X-ray emission from early-type galaxies  \citep{Sullivan.et.al.2003,Fabbiano2006}. Most ETGs are known to be strong X-ray emitters. The contribution from various sources (like hot gas, active galactic nuclei (AGNs), supermassive black hole (SMBH) and X-ray binaries) to this X-ray emission of ETGs has been a hot topic \citep{Kim.and.Fabbiano2003,Pellegrini2010,Boroson.et.al.2011}. 

Hot gas that has been entrapped in the ISM of ETGs by the galaxy's gravitational potential is considered responsible for the X-ray emission \citep{Canizares.1999,Goulding.et.al.2016}. One of the most fundamental X-ray properties is the gas temperature ($\rm T_{GAS}$), which is also a measure of the gravitational potential. Older galaxies contain vast amounts of hot diffuse gas the temperature of which is proportional to its emission in X-rays \citep{Kim.and.Fabbiano2015,Goulding.et.al.2016,Gaspari.et.al.2019}. The studies using X-ray scaling relations on the role of hot ISM in the evolution scenario of ETGs suggested that the total galaxy mass is the regulating factor of the amount of hot gas \citep{Mathews2003,Boroson.et.al.2011,Kim.and.Fabbiano2013,Civano.et.al2014}. Numerical simulations also used to reproduce the X-ray scaling relations and verified that X-ray gas luminosity (\lx), and temperature of the hot gas are proportional to the galactic mass \citep{Negri.et.al.2014,Choi.et.al2015}. 

The mass of the central black hole, $\rm M_{BH}$, of ETGs is well correlated with X-ray gas luminosity and gas temperature of the host galaxy. No optical variable, including stellar velocity dispersion ($\sigma$), appears to have stronger correlations than the X-ray gas luminosity when considering intrinsic scatter and correlation coefficient \citep{Gaspari.et.al.2019}. The strong correlations $\rm M_{BH}-T_{GAS}$ and $\rm M_{BH}-L_{X,GAS}$ imply that X-ray properties are more fundamental than optical properties \citep{Gaspari.et.al2017,Gaspari.et.al.2019}. Furthermore, the evolution of massive ETGs has been suggested using a variety of scaling relations based on X-ray luminosity with K-band magnitude ($\rm M_K$) reveals massive galaxies possess more stellar mass than the less massive systems \citep{Su.et.al2015,Kim.and.Fabbiano2015,Goulding.et.al.2016,Truong.et.al.2020}. The stellar mass loss in this systems may be the reason for the abundance of gas in the ISM, and $\rm L_{X,GAS}-Age$ relation \citep{Boroson.et.al.2011} suggests that old galaxies are massive and harbour large amount of hot gas in the ISM of host galaxies.

ETGs have been classified into two categories: core and coreless galaxies, based on their central surface brightness profile \citep{Kim.and.Fabbiano2015}. Core galaxies are considered to evolve via dry mergers with reducing star formation, while the coreless galaxies could be the outcome of gas-rich wet mergers with ensuing star formation \citep{Kormendy2009,Lauer2012,Gabor2015,Kim.and.Fabbiano2015}. The hot gas that produces X-rays is more concentrated in core galaxies \citep{Pellegrini2005,Kormendy2009,Lauer2012,Sarzi.et.al2013}, also, they have large $\rm L_K$ with no evidence of recent star formation \citep{Kormendy2009,Lauer2012}. The galaxy properties of coreless galaxies are the exact reverse of those of core galaxies. While certain core galaxies may have low concentrations of hot gas, all coreless galaxies are hot gas poor \citep{Pellegrini1999,Kim.and.Fabbiano2015}.

The central intensity ratio, CIR, found to have strong correlations with features of ETGs \citep[e.g. $\rm M_{BH}$, $\sigma$, $\rm M_{bulge}$, and $\rm M_{gal}$;][]{Aswathy2018}. Additionally, CIR provides important insights into the central star formation of host galaxies and exhibits substantial correlations with the structural parameters of late-type galaxies \citep{Aswathy2020}. The outshining of spectral lines due to AGN activity seems not to affect the estimation of CIR significantly and their use to determine the SMBH masses in Seyfert galaxies \citep{Vinod2023}. In this context, we investigate the connection between the CIR and hot gas properties of ETGs in the nearby Universe.

This paper is structured as follows. Section 2 describes the properties of the sample galaxies and the data reduction techniques employed in this study. Section 3 deals with results consisting of various correlations. Discussion and conclusion are provided in Section 4.

\section{The sample and observations}

The study involves analysis of a representative sample of nearby ($\rm D<30\, Mpc$) ETGs. The sample consists of all ETGs with measurements of X-ray luminosity from hot gas (\lx) and Hubble Space Telescope (HST) observation using the filter F814W of Wide Field Planetary Camera 2 (WFPC2). The sample consists of 27 ellipticals and  9 lenticulars. In addition,  we have included the Sombrero galaxy (NGC4594) which was classified earlier as spiral but later reported to have dual morphology \citep[see,][]{Gadotti2012}. We included this galaxy in our analysis as it is indistinguishable in all correlations exhibited by ellipticals in our sample.  Further, we have avoided the completely edge-on lenticular galaxy NGC5866, with dust lanes along the major axis of its image, as the uncertainty in its CIR value could be high. Thus the final sample consists of 36 galaxies and are listed in Table \ref{table1}.

\subsection{The Central Intensity Ratio}
Following \cite{Aswathy2018}, we computed the CIR for sample galaxies using the aperture photometry tool $MAG\_APER$  provided in the Source Extractor \citep[SExtractor,][]{Bertin.and.Arnouts1996}.

\begin{equation}
\rm CIR = \frac{I_1}{I_2-I_1}=\frac{10^{0.4(m_2-m_1)}}{1-10^{0.4(m_2-m_1)}}
\end{equation}

where $\rm I_1$ and $\rm I_2$ are the intensities and $\rm m_1$ and $\rm m_2$ are the corresponding 
magnitudes of the light within the inner ($\rm r_1=1.5$ arcsecs) and outer ($\rm r_2=3$ arcsecs) apertures, respectively. 
The simple definition of CIR makes it independent of the central intensity, I(0), for galaxies with surface brightness at a radial distance r, I(r) = I(0)f(r), where f(r) is a function of r. Secondly, the definition boosts any addition to (or subtraction from) the central intensity I(0). Thirdly, CIR, being a ratio involving only surface brightness, shows remarkable stability over a range of radii (with $r_2 = 2r_1$) despite variations in distance and orientation of the systems, which can be validated using simple Monte Carlo simulations \citep{Aswathy2018,Aswathy2020}.

Star formation, dust absorption and AGN activity in host galaxies are some of the factors that can affect the estimation of CIR. However, issues related to star formation and dust absorption are expected to be minimal in the sample of ETGs. In order to assess the effect of AGN activity in the optical images, we examined the central regions of the residual images by constructing model images using the task {\it ellipse} in IRAF.  None of our sample galaxies show an extreme central excess in optical intensity that could affect the estimation of CIR, except one, NGC4486, which has strong jet emission in its central region and is an outlier in almost all our correlations.

\begin{table*}
    \addtolength{\tabcolsep}{0.5pt}
	\centering
	\caption{Galaxy sample and properties }
	\vspace{-2mm}
	\begin{tabular}{lccccclllllc} 
	\hline
	Galaxy & Distance & $\rm log\, M_{*}$ & Morphology & $\rm CIR$ & $\Delta$\rm CIR & \rm log\, L$_{\rm X,GAS}$ & \rm T$_{\rm GAS}$ & Age & $\rm M_K$  & $\rm log\, M_{BH}$ & Profile\\
 &(Mpc) & $(\rm M_{\odot})$ & &  & &$(\rm erg \, s^{-1})$ & (keV) & (Gyr) & (mag) & $(\rm M_{\odot})$ & \\
 (1)& (2)& (3)& (4)& (5)& (6)& (7)& (8)& (9)& (10)& (11)& (12)\\
 \hline
 NGC 0524& 23.3& 10.66 & S0 & 0.82 & 0.03& 40.04$^{\dagger}$ & 0.50& 12.2$^{a}$& -24.71 & 8.94 & $\cap$\\
 NGC 0821& 23.4& 11.00 & E & 1.23 & 0.06& 38.40& 0.09& 11.0& -23.99& 7.59 & $\wedge$\\
 NGC 1023& 11.1& 10.99 & S0 & 1.03 & 0.02& 38.79& 0.30& 12.3& -24.01& 7.62 & $\wedge$\\
 NGC 1316& 20.8& 11.55 & S0 & 1.11 & 0.01& 40.70& - & 4.7& -26.00$^e$ & 8.18 & $\cap$\\
 NGC 1332& 21.5& 10.67 & E & 0.88 & 0.02& 39.90$^{\dagger}$& 0.62$^{\dagger}$& - & -24.61$^e$ & 8.83 & $\cap$\\
 NGC 1374& 18.1& 10.52 & E & 1.27 & 0.03& 38.60$^{\dagger}$& 0.86$^{\dagger}$& - & -23.10$^e$ & 8.74 & -\\
 NGC 1399& 21.2& 11.50 & E & 0.58 & 0.01& 41.73& 1.01$^{\dagger}$& 11.0 & -25.31$^e$ & 8.94 & $\cap$\\
 NGC 1400& 26.8& 11.08 & E/S0& 1.08 & 0.02& 39.67& - & 9.57$^d$ & -24.31$^e$ & 8.04$^g$ & $\wedge$\\
 NGC 2768& 21.8& 11.21 & E/S0& 1.01 & 0.02& 39.88& 0.31& 12.3& -24.71& 8.41$^g$ & $\wedge$\\
 NGC 2974& 20.9& 10.93 & E & 1.20 & 0.04& 39.30& - & 9.3 & -23.62& 8.23$^f$ & $\wedge$\\
 NGC 3115& 9.4& 10.93 & S0 & 0.82 & 0.01& 40.40& - & 9.0 & -23.99$^e$ & 8.95$^g$ & $\wedge$\\
 NGC 3377& 10.9& 10.50 & E & 1.41 & 0.04& 38.00& 0.19& 7.0 & -22.76& 7.89 & $\wedge$\\
 NGC 3585& 17.3& 10.71 & E & 0.97 & 0.02& 39.36$^{\dagger}$& 0.31$^{\dagger}$& - & -24.50$^e$ & 8.52 & -\\
 NGC 3607& 22.2& 11.39 & S0 & 0.67 & 0.02& 40.20& 0.59& 10.3& -24.74& 8.14 & $\cap$\\
 NGC 3608& 22.3& 11.03 & E & 1.21 & 0.06& 39.62& 0.40& 9.9& -23.65& 8.32 & $\cap$\\
 NGC 4261& 29.4& 11.46 & E & 0.69 & 0.02& 40.82$^{\ast}$& 0.76& 16.3$^a$ & -25.18& 8.72 & $\cap$\\
 NGC 4278& 15.6& 10.95 & E & 0.89 & 0.03& 39.39& 0.30& 11.8& -23.80& 7.96 & $\cap$\\
 NGC 4365& 23.3& 11.51 & E & 0.69 & 0.02& 39.66& 0.46& 13.4& -25.21& 8.85$^i$ & $\cap$\\
 NGC 4374& 18.5& 11.51 & E & 0.60 & 0.01& 40.82& 0.73& 13.7$^b$ & -25.12& 8.97 & $\cap$\\
 NGC 4382& 17.9& 10.90 & S0-a & 0.92 & 0.02& 39.80$^{\ast}$& 0.39& 1.6$^a$ & -25.13& 7.11$^f$ & $\cap$\\
 NGC 4406& 16.8& 11.60 & E & 0.85 & 0.02& 41.01$^{\ast}$& 0.82& - & -25.04& 8.06$^i$ & $\cap$\\
 NGC 4459& 16.1& 10.98 & S0 & 1.26 & 0.04& 39.39& 0.40& 7.0& -23.89 & 7.84 & $\wedge$\\
 NGC 4472& 17.1& 11.78 & E & 0.53 & 0.02& 41.36& 0.95& 10.5$^c$ & -25.78 & 9.40 & $\cap$\\
 NGC 4473& 15.3& 10.96 & E & 0.86 & 0.01& 39.09& 0.31& 13.1& -23.77& 7.95 & $\cap$\\
 NGC 4486& 17.2& 11.62 & E & 0.53 & 0.02& 42.93& 1.50& 12.7$^b$ & -25.38& 9.81 & $\cap$\\
 NGC 4494& 16.6& 11.02 & E & 1.14 & 0.03& 39.11& 0.34& 8.0& -24.11& 7.74$^h$ & $\wedge$\\
 NGC 4526& 16.4& 11.26 & S0 & 0.98 & 0.03& 39.45& 0.31& 11.0& -24.62& 8.65 & -\\
 NGC 4552& 15.8& 10.87 & E & 0.91 & 0.02& 40.30$^{\dagger}$& 0.59& 12.4$^a$ & -24.29& 8.70 & $\cap$\\
 NGC 4594& 9.77& 11.41 & Sa & 0.91 & 0.01& 39.32& - & 12.5$^b$ & -24.99$^e$ & 8.82 & $\cap$\\
 NGC 4621& 14.9& 10.92 & E & 1.24 & 0.02& 38.48$^{\dagger}$& 0.27& 10.6$^c$ & -24.14& 8.59 & $\wedge$\\
 NGC 4636& 14.3& 11.17 & E & 0.57 & 0.02& 41.50& 0.73& 13.4& -24.36 & 8.58 & $\cap$\\
 NGC 4649& 17.3& 11.60 & E/S0 & 0.50 & 0.01& 41.22& 0.86& 13.2$^b$ & -25.46& 9.32 & $\cap$\\
 NGC 5128& 5.8& 10.94 & E/S0 & 0.48 & 0.03& 40.20& - & - & -24.89$^e$ & 7.76 & -\\
 NGC 5576& 24.8& 11.04 & E & 1.13 & 0.05& 38.95$^{\dagger}$& 0.52& 2.5$^a$ & -24.15& 8.23 & $\wedge$\\
 NGC 5846& 24.2& 11.46 & E/S0 & 0.66 & 0.02& 41.70& 0.72& 12.7$^b$ & -25.01& 9.04 & $\cap$\\
 NGC 7457& 12.9& 10.13 & S0 & 1.20 & 0.07& 38.08& 0.30& 3.8& -22.38& 6.95$^f$ & $\wedge$\\

		\hline	
	\end{tabular}
	\begin{flushleft}
	\vspace{-2mm}
	\footnotesize
Notes: Columns are (1) Galaxy name; (2) Distance; (3) Stellar mass; (4) Morphology; (5) Central Intensity Ratio (CIR); (6) uncertainty of CIR; (7) X-ray gas luminosity in $0.3-8$ keV band from \cite{DuncanForbes.2017} [except $\dagger =$ \cite{Lakhchaura2019}, $\ast =$ \cite{Goulding.et.al.2016}]; (8) Gas temperature from \cite{Kim.and.Fabbiano2015} [except $\dagger =$ \cite{Lakhchaura2019}]; (9) Age of the central stellar population from \cite{Alabi.et.al.2017} [except a = \cite{Kim.and.Fabbiano2015}, b = \cite{DuncanForbes.2017a}, c = \cite{Yamada.et.al2006}, d = \cite{Barr.et.al2007}]; (10) K-band magnitude from \cite{Cappellari.et.al.2011} [except e = derived from the apparent magnitude $K_T$ (2MASS keyword {k{\_}m{\_}ext}) using the method adopted by \cite{Cappellari.et.al.2011}; (11) Mass of the SMBH from \cite{Gaspari.et.al.2019} [except f = \cite{vandenBosch2016}, g = \cite{Pota2013}, h = \cite{Pellegrini2010}, i = \cite{vanderMarel1999}]; (12) Central light profile: $\cap$ = core, $\wedge$ = coreless from \cite{Krajnovic.et.al2013} and \cite{Pellegrini1999}.
\end{flushleft}
\label{table1}	
\end{table*}

\begin{figure*}
    \centering
    \begin{subfigure}[h!]{0.45\linewidth}
        \includegraphics[height=70mm,width=95mm,angle=0]{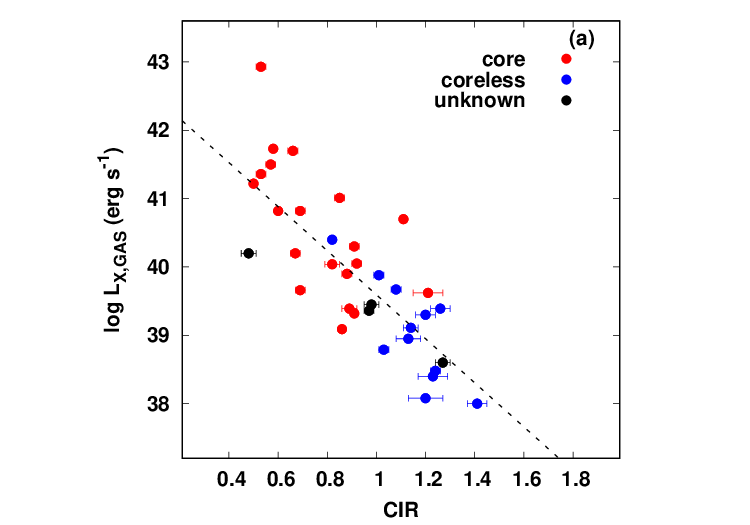}
    \end{subfigure}
        \vspace{4mm}
        \hspace*{2mm}
    \begin{subfigure}[h!]{0.45\linewidth}
        \includegraphics[height=70mm,width=95mm,angle=0]{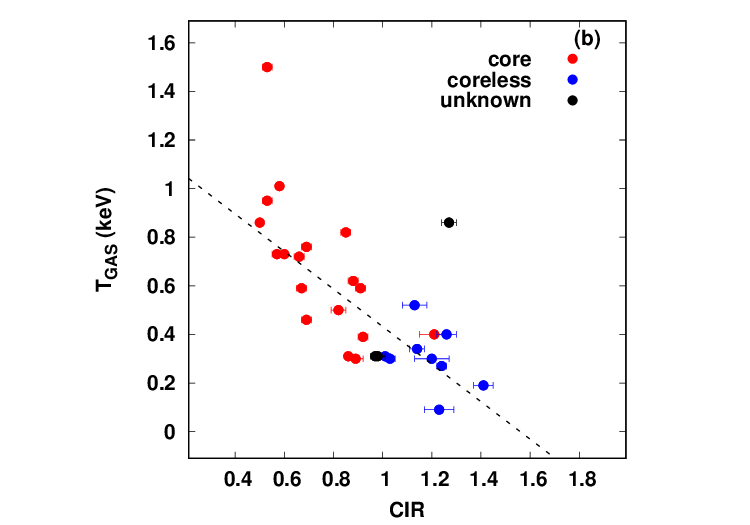}
    \end{subfigure}
    \vspace{4mm}
    \hspace*{2mm}
    \centering
    \hspace*{2mm}
    \begin{subfigure}[h!]{0.45\linewidth}
        \includegraphics[height=70mm,width=95mm,angle=0]{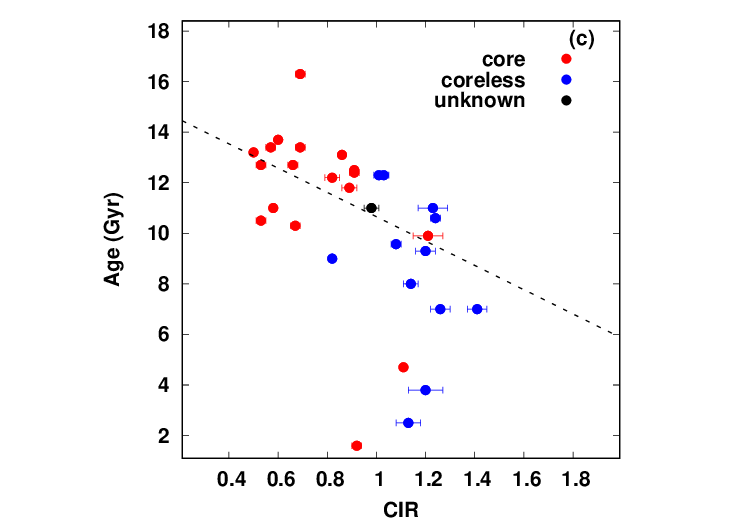}
    \end{subfigure}
        \vspace{4mm}
        \hspace*{3mm}
    \begin{subfigure}[h!]{0.45\linewidth}
        \includegraphics[height=70mm,width=95mm,angle=0]{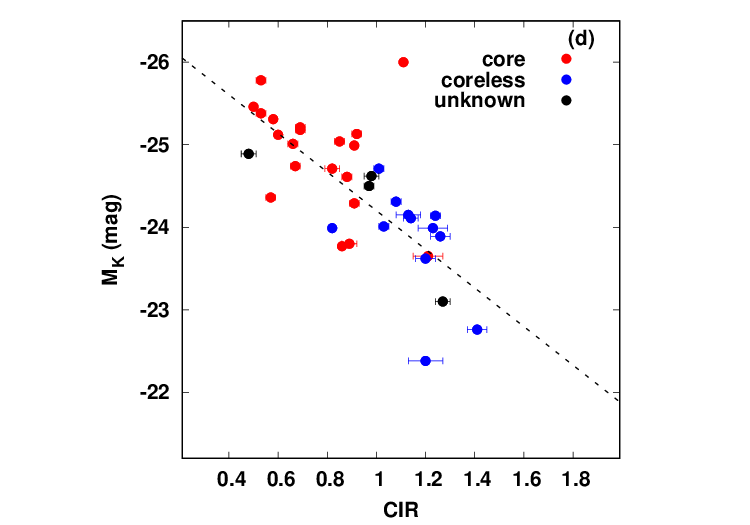}
    \end{subfigure}
    \hspace*{5mm}
        \caption{Variation between the central intensity ratio and (a) X-ray gas luminosity (b) Gas temperature (c) Age of central stellar population and (d) K-band magnitude of the sample galaxies. The dashed line indicates the best fitting.}
        \label{fig1}
\end{figure*}

\section{RESULTS}
The estimated CIR of ETGs is well correlated with the properties of hot diffuse gas in the host galaxy. The scaling relations obtained between CIR and gas properties of galaxies shed some light on the influence of hot gas in the evolution scenario of ETGs. The estimated CIR values and other properties of sample galaxies are listed in Table \ref{table1}. As seen in Table \ref{table1} different properties of galaxies are assembled from various authors. Estimations of X-ray luminosity from hot gas in the 0.3 - 8 keV band were taken from \cite{DuncanForbes.2017}, \cite{Lakhchaura2019}, and \cite{Goulding.et.al.2016}. The temperature of the hot gas content was 
 compiled from \cite{Kim.and.Fabbiano2015} and \cite{Lakhchaura2019}. K-band magnitude of the galaxies determined using 2MASS observations were taken from \cite{Cappellari.et.al.2011} and we derived K-band magnitude for the sources which are not included by \cite{Cappellari.et.al.2011}. Core or coreless characterization of the central light profile of the sample galaxies was taken from \cite{Krajnovic.et.al2013} and \cite{Pellegrini1999}. 

\subsection{Correlation between the CIR and L$_{\rm X,GAS}$}
We find a strong correlation between the CIR and X-ray gas luminosity of early-type galaxies. The observed correlation between CIR and L$_{\rm X,GAS}$ is shown in Fig. \ref{fig1}a. The linear correlation coefficient of this correlation, without considering errors, is $r=-0.79$ with the null hypothesis probability, $p = 0.0001$  \citep{Press.et.al.1992}. The best-fit parameters of a linear fit along with the correlation coefficients are given in Table \ref{table2}.
ETGs follow strong positive correlation between L$_{\rm X,GAS}$ and T$_{\rm GAS}$ \citep{Sullivan.et.al.2003,Boroson.et.al.2011,Goulding.et.al.2016}. The observed negative correlation between CIR and L$_{\rm X,GAS}$, on the other hand, possibly reflects the suppression in the central optical light as L$_{\rm X,GAS}$ (and T$_{\rm GAS}$) increases. Due to the high temperature of diffuse gas, the star formation might be quenched in the host galaxy. The SMBH mass and T$_{\rm GAS}$ of ETGs are found to be positively correlated \citep{Gaspari.et.al.2019}. The strength of feedback might be proportional to $\rm M_{BH}$ and the gas temperature can increase near the vicinity of SMBH, suggesting the quenching of star formation will be more in the central region than in the outskirts of the host galaxy. This quenching of star formation can reduce the value of optical CIR. 

The galaxy NGC4486 is an outlier in Fig. \ref{fig1}a. It is a massive elliptical galaxy that emits a synchrotron jet from its non-thermal core. The jet morphology of NGC4486 has been reported to be present at several wavelengths ranging from radio to X-ray \citep[e.g.,][]{Marshall.et.al.2002,Xilouris.et.al.2004,Ferrarese.et.al.2006}. The jet phenomenon emerging from the core of this galaxy can expel hot gas from the central region. The feedback mechanism from the central region of galaxies may suppress star formation and decrease the CIR value. NGC4486 is the central galaxy of the Virgo cluster and is surrounded by enormously hot and diffuse gas, indicating its deep potential well \citep{Kormendy2009}. During the gravitational assembly of clusters or groups of galaxies, the temperature of this hot gas reaches up to $10^{7 - 8}$K \citep{Mernier2017}. The presence of this extreme hot intra-cluster medium (ICM) prevalent in the galaxy, which is responsible for the excess X-ray emission, however, is not effecting a change in CIR, probably because the feedback here is spread much above the inner aperture considered. 

We identified that there is a distinction between core and coreless galaxies in the $\rm CIR-L_{\rm X,GAS}$ relation, with the core galaxies having larger L$_{\rm X,GAS}$ and lower CIR values, while coreless galaxies exhibit the reverse (see Fig. \ref{fig1}). The core galaxies are believed to be formed through dry mergers with suppressed star formation, while wet mergers coupled with enhanced star formation result in coreless galaxies \citep{Kim.and.Fabbiano2015}. So, the $\rm CIR-L_{\rm X,GAS}$ relation suggests that the star formation process might have been quenched due to the lack of cold gas in core galaxies. Two galaxies (NGC3608 and NGC3115) appear to be a misfit in the core-coreless classification. NGC3608 ($ \rm CIR=1.21$) has a definite core profile, and has been reported as a recent merger \citep{Pinkney2003}. The merger event could increase the star formation of the galaxy and affect the value of CIR\footnote{The reduced L$_{\rm X,GAS}$ value of this galaxy might also be connected to the recent star formation \citep{Boroson.et.al.2011}.}. The other galaxy, NGC3115 ($ \rm CIR=0.82$), which is defined as a coreless galaxy, is a lenticular with a prominent disk. The disk is viewed nearly edge-on, and the inclination effect renders its CIR value unreliable.

\subsection{Correlation between the CIR and T$_{\rm GAS}$}
We observed a strong correlation between the CIR and temperature of the diffused gas inside the galaxy with a correlation coefficient, $r=-0.78$ with $p = 0.0001$ (Fig. \ref{fig1}b). The connection between CIR and T$_{\rm GAS}$ is apparent, as the latter strongly correlates with X-ray gas luminosity. It is commonly known that X-ray luminosity and gas temperature are positively correlated in massive galaxies \citep{Sullivan.et.al.2003,Mathews2003,Boroson.et.al.2011,Negri.et.al.2014}. Our analysis further supported this argument, as we observed that the CIR is low when the gas temperature is high. The high temperature prevents star formation and lowers optical emission in the core areas of host galaxies. 

At the same time, two galaxies, NGC4486 and NGC1374, deviated from the best fit of the $\rm CIR-T_{\rm GAS}$ correlation. NGC4486 is also an outlier in the $\rm CIR-L_{\rm X,GAS}$ relationship, with the peculiarities and reasons for its deviation from the fit described previously (see section 3.1). Galaxy NGC1374 is reported to have high T$_{\rm GAS}$ ($\rm 0.86 \,keV$) when compared with its L$_{\rm X,GAS}$ ($\rm 3.98 \times 10^{38}\, erg \, s^{-1}$) and could be a gas-poor ETG \citep{Kim.and.Fabbiano2015}.

The CIR of ETGs shows significant correlation (\mbox{$r = -0.66$,} $p = 0.0002$) with the age of the central stellar population within R$_e$ (Fig. \ref{fig1}c) barring a few galaxies younger than 6 Gyr in our sample. Given the uncertainties involved in the estimation of the age of galaxies using single stellar population models \citep{Conroy2010,Raichoor.et.al.2011}, it is remarkable that the simple photometric parameter, CIR, successfully represents the age of ETGs.  The $\rm CIR-Age$ correlation, can be a restatement of correlation between L$_{\rm X,GAS}$ and Age \citep{Boroson.et.al.2011}. However, the galaxies younger than 6 Gyr seem to deviate from the observed $\rm CIR-Age$ correlation, and possibly suggests an increased uncertainty in the estimation of age there.

K-band absolute magnitude $\rm (M_K)$, one of the proxies for stellar mass, displays strong association with CIR ($r = 0.79$ with $p = 0.0001$) and is shown in  Fig. \ref{fig1}d. The L$_{\rm X,GAS}$ and T$_{\rm GAS}$ of nearby ETGs display strong correlations with $\rm M_K$ in both observations as well as simulations \citep{Truong.et.al.2020}. The existence of $\rm M_{BH}-M_K$ correlation is also known \citep{Graham2013}. The pure photometric strong $\rm CIR-M_K$ correlation, which may be considered as reflection of the influence of the central SMBHs in ETGs, is the strongest when compared with L$_{\rm X,GAS}- \rm M_K$, T$_{\rm GAS}- \rm M_K$ and $\rm M_{BH}-M_K$ in terms of correlation coefficient and scatter.

\begin{table}
\addtolength{\tabcolsep}{-2pt}
\caption{The table lists the best-fitting parameters for the relation $x = \alpha\rm CIR + \beta$ and correlation coefficients for various relations.}          
\begin{tabular}{l c c c c c }
\hline
 x	&	$\alpha$	& $\beta$	&	r	&	p	&	n \\
\hline
log $\rm L_{X,GAS}$ &  -3.22$\pm$0.41 & \ 42.81$\pm$0.39 & -0.79 & $  0.0001$ & 35\\
log $\rm T_{GAS}$ &  -0.77$\pm$0.11 & \ \ \,1.20$\pm$0.10 & -0.78 & $  0.0001$ & 28\\
Age &  -4.81$\pm$1.09 & \ 15.46$\pm$1.02 & -0.66 & $  0.0002$ & 26\\
$\rm M_{K}$ &  \ 2.34$\pm$0.31 & -26.54$\pm$0.29 & \ 0.79 & $  0.0001$ & 35\\
\hline
\end{tabular}
\label{table2}
\end{table}

\section{Discussion and Conclusion}
The central region of galaxies plays a major role in their evolution mechanism. ETGs have a thermal ISM that produces X-ray gas luminosity and temperature, which are perhaps the essential measurables in such systems and follow a universal scaling relation of L$_{\rm X,GAS}$ $\propto$ T$_{\rm GAS}$, enhancing understanding of the connection between gas temperature and galaxy potential \citep{Goulding.et.al.2016}. In the study of groups and clusters of galaxies, the L$_{\rm X,GAS}-\rm T_{\rm GAS}$ relationship has been widely used to perceive the evolution of host galaxies \citep{Sullivan.et.al.2003}. In ETGs, a single power-law relationship between L$_{\rm X,GAS}$ and T$_{\rm GAS}$ explains the hot gas distribution within the core region of the host galaxies, where the gas temperature governs the observed X-ray gas luminosity.

 The existence of core and coreless galaxies in ETGs is already studied in detail in the literature \citep{Kormendy2009ASPC,Krajnovic.et.al2013,Kim.and.Fabbiano2015}. Core galaxies are defined primarily by dry mergers with restrained star formation while coreless galaxies are considered to be formed through gas-rich wet mergers \citep{Kormendy2009,Lauer2012,Gabor2015}. Core galaxies possess lower CIR values in comparison with coreless galaxies. The simple photometric parameter CIR is capable of distinguishing between core and coreless ETGs. More importantly, our study shows that the CIR measures the steepness of the central intensity profile so naturally that the dichotomy is translated into a continuous variation in CIR as more and more ETGs turn coreless as CIR increases. 

For the first time, we identify significant correlations between CIR and X-ray properties of ETGs like,  L$_{\rm X,GAS}$, and T$_{\rm GAS}$ (Figures \ref{fig1}a and \ref{fig1}b). The thermalization of the energy from stellar mass loss material caused by evolved stars often produces the X-ray luminosity in ETGs \citep{Goulding.et.al.2016}. Optical CIR carries information about the star formation in the central region of ETGs, and older galaxies possess lower CIR values \citep{Aswathy2018}. The decrease in CIR indicates that star formation is much less near the centre of galaxies than in the outskirts. The hot diffuse gas in the ISM of ETGs may drive the stellar population and regulate the optical emission from the central region of the galaxies. Moreover, the flux ratio X-ray/optical seems to decrease systematically from the nuclear region \citep{Marshall.et.al.2002} of host galaxies. The amount of hot diffuse gas in older galaxies is higher \citep{Boroson.et.al.2011}, and this hot environment may prevent star formation in the central region, lowering the CIR value.

The strong correlations exhibited by L$_{\rm X,GAS}$ and T$_{\rm GAS}$ independently with CIR  can be used for their estimation. Currently, the total mass of a galaxy ($\rm M_{total}$) including dark matter, is the only quantity that can be used for the estimation of L$_{\rm X,GAS}$. However, disentangling stellar content and dark matter halo is a challenging process while doing dynamical modeling to estimate  $\rm M_{total}$ \citep{Kim.and.Fabbiano2013}. For the ETGs in this study, the scatter obtained for the relations $\rm M_{total}-L_{\rm X,GAS}$ and $\rm CIR-L_{\rm X,GAS}$, are 0.72 and 0.73 dex, respectively. We find that the CIR not only shows continuous scaling relations but also has the remarkable potential to estimate  L$_{\rm X,GAS}$. Similarly, the small scatter in the $\rm CIR-T_{\rm GAS}$ relation is -0.86 dex and is useful in the quick estimation of the latter. 

The hot diffuse gas confined in the potential well of ETGs is believed to be emitting X-rays \citep{Canizares.1999} and core elliptical galaxies are reported to host a greater abundance of hot gas than coreless galaxies \citep{Kim.and.Fabbiano2015}. 
\cite{Kormendy2009} suggested that the hot ISM of core ETGs may provide the essential environment for a feedback process that prevents gas accretion and impedes star formation. The feedback mechanism of the host galaxy may have a major impact on the suppression or prevention of star formation \citep{Morganti2017}. The central engine of a galaxy, the SMBH, also plays a crucial role in preventing star formation by expelling gas from the central region of the host galaxy \citep{Werner.et.al.2014}. 
The continuous variation of CIR with hot gas properties of ETGs, along with its reported correlation with MBH, on the other hand, appears to reflect its potential to trace the effect of feedback driven by the central SMBH.  As $\rm M_{BH}$ grows, the abundance of hot gas also rises, producing significant  X-ray emission \citep{Gaspari.et.al.2019}.  
 At the same time, the optical light in the inner region of galaxies may be suppressed more as the SMBH mass increases \citep{Aswathy2018,Vinod2023}. Thus, the observed  $\rm CIR-T_{\rm GAS}$ anti-correlation supports a scenario where an increase in the (local) gas temperature results in increased suppression of optical intensity. Hence, while $\rm T_{\rm GAS}$ represents a kind of average gas temperature within R$_e$, the CIR seems to reflect the effect of local gas temperature on star formation at the central region in ETGs.

 ETGs display characteristic dichotomy in many of their properties. The E-E dichotomy brought out the core, coreless classification \citep{Kormendy2009ASPC, Kormendy2009}. In addition to this, the existence of boxy-disky, core-cusp, and slow and fast rotating galaxies are discussed in the literature \citep{Kormendy2009, Boroson.et.al.2011, Kim.and.Fabbiano2015}.  L$_{\rm X,GAS}$, and T$_{\rm GAS}$ are higher for core galaxies than coreless \citep{Kim.and.Fabbiano2015}, while this study, shows CIR has the potential to convert this dichotomy into a gradual sequence.  By its definition, CIR is designed to boost any difference in intensity existing in the surface brightness profile of a galaxy from a smooth function. Any addition of intensity near the centre of a galaxy, say by an enhanced star formation, can increase the CIR, while suppression of light in the outer regions, say by dust extinction,  can also produce similar effect. Hence one has to carefully select a sample, so that CIR is put to maximum use. The many correlations that CIR exhibits with galaxy properties demonstrate that it has great potential in understanding galaxy evolution.

\section*{Acknowledgements}

We sincerely thank the anonymous referee for her/his comments which had improved the quality of the paper significantly. VKT would like to acknowledge the financial support from Council of Scientific \& Industrial Research (CSIR), Government of India. VKT is thankful to UGC-SAP and FIST 2 (SR/FIST/PS1-159/2010) (DST, Government of India) for the research facilities in the Department of Physics, University of Calicut. We acknowledge the use of the NASA/IPAC Extragalactic Database (NED), \url{https://ned.ipac.caltech.edu/} operated by the Jet Propulsion Laboratory, California Institute of Technology, and the Hyperleda database, \url{http://leda.univ-lyon1.fr/}. We acknowledge the use of data publicly available at Mikulski Archive for Space Telescopes (MAST), \url{http://archive.stsci.edu/} observed by NASA/ESA Hubble Space Telescope (HST).

\section*{Data Availability}

The data adopted for this paper are publicly available in FITS format at Mikulski Archive for Space Telescopes (MAST), \url{http://archive.stsci.edu/} at the Space Telescope Science Institute (STScI) observed by the NASA/ESA Hubble Space Telescope (HST).



\bibliographystyle{mnras}
\bibliography{reference.bib} 






\bsp	
\label{lastpage}
\end{document}